\UseRawInputEncoding
\documentclass[%
 reprint,
 amsmath,amssymb,
 aps,
superscriptaddress
]{revtex4-1}
\usepackage{filecontents}
\usepackage{graphicx}
\usepackage{dcolumn}
\usepackage{bm}
\usepackage[dvipsnames]{xcolor}
\usepackage{mathtools}
\usepackage[utf8]{inputenc}
\usepackage{csquotes}
\usepackage{ulem} 
\usepackage[english]{babel}
\usepackage{textcomp}
\usepackage{amstext} 
\usepackage{array} 
\usepackage{siunitx}
\usepackage{cancel}

\newcommand\eff{\textrm{eff}}
\newcommand\A{\textrm{A}}

\newcommand\CW{\textrm{cw}}
\newcommand\CCW{\textrm{ccw}}
\newcommand\inn{\textrm{in}}
\newcommand\out{\textrm{out}}
\addto\captionsenglish{}

\begin{document}

\title{Kerr-Nonlinearity-Induced Mode-Splitting in Optical Microresonators}

\author{George~N.~Ghalanos}
\affiliation{Max Planck Institute for the Science of Light, Staudtstraße 2, 91058 Erlangen, Germany}
\affiliation{National Physical Laboratory, Hampton Road, Teddington, TW11 0LW, UK}
\affiliation{Blackett Laboratory, Imperial College London, SW7 2AZ, UK}
\author{Jonathan~M.~Silver}
\affiliation{National Physical Laboratory, Hampton Road, Teddington, TW11 0LW, UK}
\affiliation{City, University of London, EC1V 0HB, UK}
\author{Leonardo~Del~Bino}
\affiliation{National Physical Laboratory, Hampton Road, Teddington, TW11 0LW, UK}
\affiliation{Heriot-Watt University, Edinburgh, EH14 4AS, UK}
\author{Niall~Moroney}
\affiliation{Max Planck Institute for the Science of Light, Staudtstraße 2, 91058 Erlangen, Germany}
\affiliation{National Physical Laboratory, Hampton Road, Teddington, TW11 0LW, UK}
\affiliation{Blackett Laboratory, Imperial College London, SW7 2AZ, UK}
\author{Shuangyou~Zhang}
\affiliation{Max Planck Institute for the Science of Light, Staudtstraße 2, 91058 Erlangen, Germany}
\affiliation{National Physical Laboratory, Hampton Road, Teddington, TW11 0LW, UK}
\author{Michael~T.~M.~Woodley}
\affiliation{National Physical Laboratory, Hampton Road, Teddington, TW11 0LW, UK}
\affiliation{Heriot-Watt University, Edinburgh, EH14 4AS, UK}
\author{Andreas~\O.~Svela}
\affiliation{National Physical Laboratory, Hampton Road, Teddington, TW11 0LW, UK}
\affiliation{Blackett Laboratory, Imperial College London, SW7 2AZ, UK}
\author{Pascal~Del'Haye}
\affiliation{Max Planck Institute for the Science of Light, Staudtstraße 2, 91058 Erlangen, Germany}
\affiliation{Department of Physics, Friedrich Alexander University Erlangen-Nuremberg, 91058 Erlangen, Germany}
\affiliation{National Physical Laboratory, Hampton Road, Teddington, TW11 0LW, UK}

\begin{abstract}

The Kerr effect in optical microresonators plays an important role for integrated photonic devices and enables third harmonic generation, four-wave mixing, and the generation of microresonator-based frequency combs. Here we experimentally demonstrate that the Kerr nonlinearity can split ultra-high-Q microresonator resonances for two continuous-wave lasers. The resonance splitting is induced by self- and cross-phase modulation and counter-intuitively enables two lasers at different wavelengths to be simultaneously resonant in the same microresonator mode. We develop a pump-probe spectroscopy scheme that allows us to measure power dependent resonance splittings of up to 35 cavity linewidths (corresponding to 52 MHz) at 10 mW of pump power. The required power to split the resonance by one cavity linewidth is only \SI{286}{\micro\watt}. In addition, we demonstrate threefold resonance splitting when taking into account four-wave mixing and two counterpropagating probe lasers. These Kerr splittings are of interest for applications that require two resonances at optically controlled offsets, eg. for opto-mechanical coupling to phonon modes, optical memories, and precisely adjustable spectral filters.

\end{abstract}

\maketitle

Whispering gallery mode (WGM) microresonators have gained much attention in recent years for their wide range of applications, particularly for optical frequency combs \cite{hansch2006nobel, del2007optical}, near-field sensing \cite{foreman2015whispering, heylman2017optical}, cavity optomechanics \cite{ruesink2016nonreciprocity, enzian2019observation}, PT symmetric systems \cite{peng2014parity}, and interaction between counterpropagating light \cite{del2017symmetry, Cao2017, Woodley2018}. The latter effect relies on the interplay between self- and cross-phase modulation (SPM and XPM respectively) in counter-propagating modes. Effects of $\chi^{(3)}$ nonlinearity, including SPM, XPM, and four-wave mixing (FWM), play an important role in many areas of nonlinear photonics and ultrafast optics. An example where all three play a role is in parametric sideband generation \cite{kippenberg2004kerr, savchenkov2004low}. The Kerr effect has been further utilised to realise optical isolators and circulators \cite{del2018microresonator}, switching \cite{Haelterman1991,Daniel2012,DelBino:21}, logic gates \cite{moroney2020logic}, gyroscopes \cite{Kaplan1981,Wang2014a,silver2020nonlinear} and near-field sensors~\cite{Wang2015}. Orthogonally polarised dual microcomb generation based on XPM has also been reported \cite{bao2019orthogonally,suzuki2018theoretical,zhang2020spectral2}.

In this work we show direct measurements demonstrating that SPM and XPM split resonances at high-circulating powers in microresonators. More specifically, two continuous wave lasers at different frequencies can be simultaneously resonant in the same resonator mode with a frequency splitting that depends on the power difference between the two lasers. We directly measure the difference between the resonance shifts seen by a strong pump and a weak probe beam that is either copropagating or counterpropagating with the pump. Moreover, in the copropagating measurement, we observe a modified SPM-XPM mode splitting due to FWM contributions. To our knowledge, these are the first spectroscopic measurements of Kerr-effect-induced two-level and three-level splittings for continuous-wave lasers in optical resonators. Importantly, we demonstrate highly resolved splittings of many cavity linewidths, made possible by the extremely high Q-factors ($>10^8$) with linewidths around 1 MHz. These fused silica microtoroids \cite{armani2003ultra} and microrod resonators \cite{del2013laser} are shown in Fig.~1(a,b). 

\begin{figure*}
 \centering\includegraphics[width=0.93\linewidth]{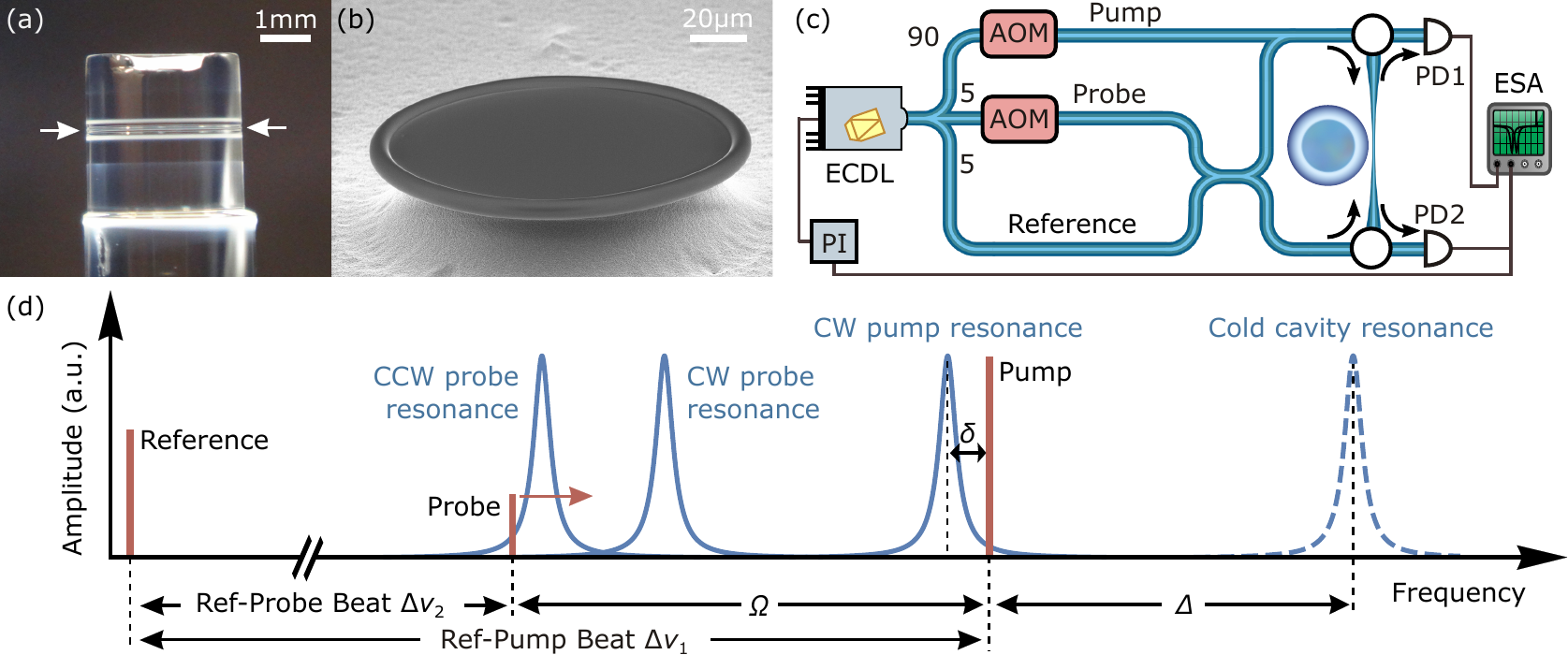}
 \caption{\label{fig:fabri1} (a) Silica rod WGM resonator with a diameter of 2.7 mm. The region in which light travels around the rod is indicated by the arrows. (b) On-chip silica microtoroid WGM resonator with a diameter of \SI{95}{\micro\meter}. (c) Experimental setup. Light from a single ECDL is split into 3 branches. The pump branch is up-shifted in frequency by a fixed amount via an AOM. A PI controller allows the pump to be locked at a fixed detuning from resonance. The probe branch is up-shifted by a variable amount via a second AOM, used for spectroscopy. The reference branch is far out of resonance and does not couple into the resonator, and is used for heterodyne detection. (d) The relative frequencies of the three beams, illustrating how the pump is locked at a fixed detuning $\delta$ from the SPM-shifted resonance, while the probe beam is scanned around the pump frequency to measure the XPM-shifted resonance. All resonances shown in solid blue originate from the same cold cavity resonance (dashed). Note that $\mathit{\Omega}$ and $\mathit{\Delta}$ are shown here as frequencies for illustrative purposes but are normalised by the cavity half-linewidth $\gamma$ in the rest of this work.
 }
\end{figure*}

The SPM-XPM splitting occurs due to power differences between the pump and probe beams. For small probe powers, the Kerr-induced frequency shifts in the absence of FWM are
\begin{equation}
\begin{split}
    \Delta\nu_{\textrm{pump}}\ &=\ \A \cdot P_{\textrm{pump}}\\
    \Delta\nu_{\textrm{probe}}\ &=\ 2\A \cdot P_{\textrm{pump}}\,,
\end{split}
\end{equation}
where $A$ is a proportionality constant. For the case in which the probe is copropagating with the pump, the factor of two between SPM-XPM can be reduced due to stimulated FWM~\cite{carman1966observation} taking place in addition to SPM and XPM. The decreased FWM-induced shift leads to a circulating-direction-dependent three-way splitting of the modes. FWM does not occur in the counterpropagating case as it would violate momentum conservation. This is also linked to recently observed resonance frequency splittings through Kerr-interaction with pulses from dissipative Kerr solitons \cite{guo2017universal}.

We use a weak probe laser to detect the splitting between SPM- and XPM-shifted resonances. To achieve this, we first split the light from a $\SI{1.55}{\micro m}$ tunable laser source into three branches, as seen in Figs.~1(c) and 1(d), before recombining them into a single tapered fibre  \cite{knight1997phase} that is coupled to a \SI{2.7}{mm} wide microrod resonator ($4\times10^8$ Q-factor) \cite{del2013laser}. The pump branch (10~-~150 mW power) is up-shifted in frequency relative to the laser by a fixed amount ($\Delta\nu_1 = $~\SI{81}{\mega\hertz}) using an acousto-optic modulator (AOM), and actively locked to a fixed detuning $\delta$ from its (thermally and SPM-shifted) resonance (see Fig.~1(d)). The transmitted pump power detected on the photodiode (PD2) is stabilised by a proportional-integral (PI) controller feeding back to the pump frequency. The active lock ensures that all other beams can be referenced relative to the pump frequency, and thus the thermally and SPM-shifted resonance. The thermally induced frequency shift affects the SPM- and XPM-resonance equally and only increases $\Delta$ in Fig.~1(d), while the microresonator thermally self-stabilizes the pump resonance to the pump laser \cite{carmon2004dynamical, ilchenko1992thermal}. 

The probe branch AOM is scanned in frequency over a range of $\pm$\SI{10}{\mega\hertz} with respect to the pump laser AOM frequency (equivalent to $\pm 4$ linewidths). The frequency offset of the probe laser is shown as $\Delta\nu_2$ in Fig.~1(d). This allows us to perform spectroscopy around the pump frequency. Unlike the pump, the probe branch is coupled to the resonator in both directions. Importantly, the probe beams are too weak to induce any significant thermal or Kerr shifts on their own. The chosen resonance is far from neighbouring resonances, ensuring that the Kerr-shifted resonance is the one probed.

The reference branch light is used for heterodyne detection, while its frequency is far away from any resonance and constantly offset by \SI{81}{\mega\hertz} from the pump laser. To measure the precise position of the XPM-shifted resonance, we electronically change $\Delta\nu_2$ to scan the probe laser across the resonance. Simultaneously, we monitor the power in the beat note between reference laser and probe laser $\Delta\nu_2$, detected by PD1 and PD2 (see Fig.~1(c)). The beat note power is reduced when the laser crosses the XPM-shifted resonance. A complete trace of the XPM-shifted resonance (see Fig.~2) is obtained by continuously measuring the beat note power with an electronic spectrum analyser in ``max hold" mode while sweeping the frequency $\Delta\nu_2$. We also ensure that the polarisation of the pump, probe and reference are aligned with each other and with the resonator modes.

\begin{figure*}
 \centering\includegraphics[width=1\linewidth]{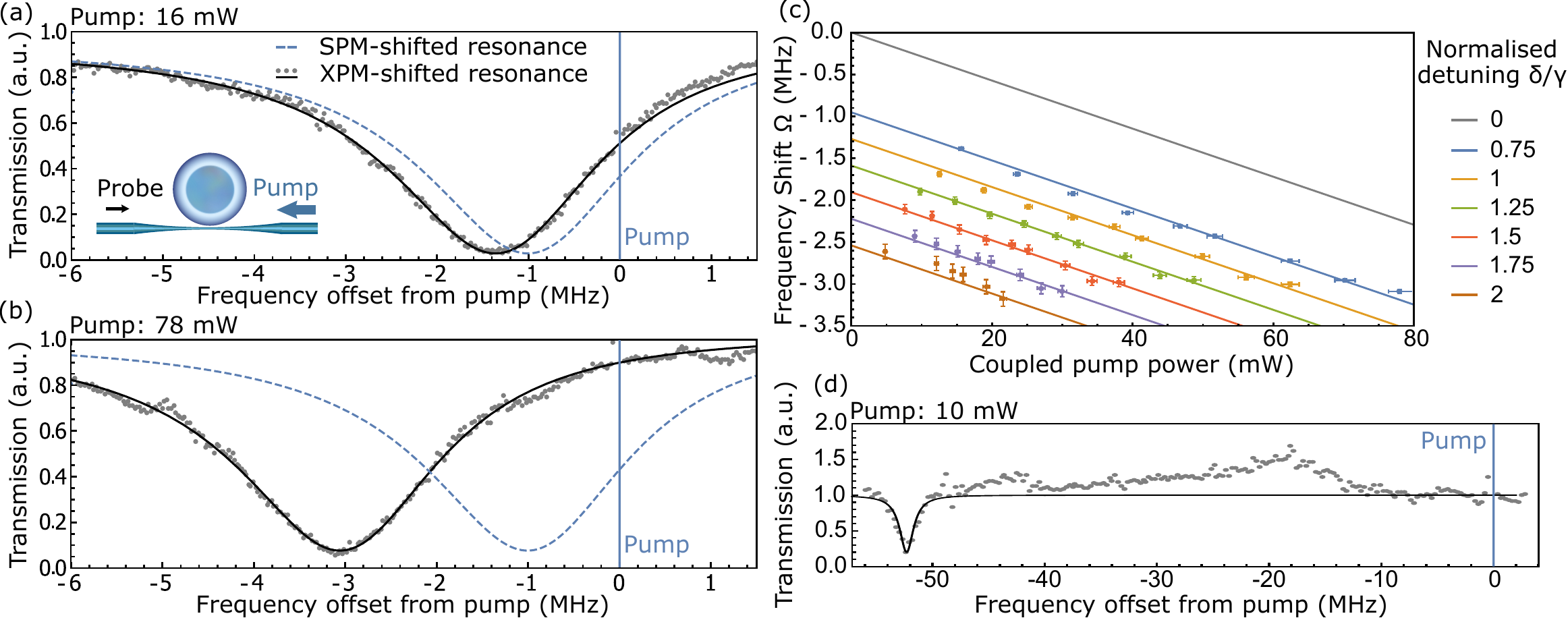}
 \caption{\label{fig:fig2} (a), (b) Probe spectra measuring the XPM shifts for different pump powers for CW-CCW pump-probe directions. With the pump locked at a fixed detuning from the SPM-shifted resonance, shown as a dashed line, the probe is scanned in frequency to measure the XPM-shifted resonance. Increasing the pump power results in a larger difference between SPM-XPM shifts. (c) XPM-related  pump-power-dependent frequency shifts. Each trace corresponds to a different detuning of the pump from the SPM resonance, in terms of half-linewidths ($\gamma$). Each data point corresponds to the frequency difference between the pump and the centre of the XPM-shifted resonance. (d) A \SI{95}{\micro\meter}-wide microtoroid resonator of $5\times10^7$ Q-factor and \SI{5}{\micro\meter^2} effective mode area $ A_{\textrm{eff}}$ was pumped with 10 mW, resulting in a difference of 70$\gamma$ between SPM-XPM shifts.
 }
\end{figure*}

The dynamics of SPM- and XPM-induced resonance shifts in a microresonator are described by the following dimensionless equations for the clockwise/counterclockwise circulating intracavity fields:
\begin{equation}
\begin{split}
    \dot{f}_{\CW} &= \tilde{f}_{\CW} - (1\!-\!i\Delta)f_{\CW} - i(|f_{\CW}|^2 + 2|f_{\CCW}|^2)f_{\CW}\\
    \dot{f}_{\CCW} &= \tilde{f}_{\CCW} \!-\! (1\!-\!i\Delta)f_{\CCW} - i(|f_{\CCW}|^2\! + \! 2|f_{\CW}|^2)f_{\CCW},
\end{split}
\end{equation}
where $\tilde{f}$ represents an external input field and where the pump and probe beams circulate in the CW and CCW directions respectively. Time is normalised by $1/\gamma$ where $\gamma$ is the cavity half-linewidth, and we are working in the rotating-wave approximation in the frame of the pump beam which is red-detuned from the cold cavity resonance by $\Delta$ (normalised by $\gamma$). Note the factor of two, arising due to XPM between the two beams. The dimensionless in-coupled powers $|f_{\CW}|^2$ and $|f_{\CCW}|^2$ are normalised by the characteristic power
\begin{equation}
    P_0 = \frac{\pi^2 n^2_0 d A_\textrm{eff}}{n_2 \lambda Q Q_0},
    \label{Pnot}
\end{equation}
where $Q$ and $Q_0$ are coupled and intrinsic Q-factors respectively, $d$ is the resonator diameter, $A_\textrm{eff}$ is the effective mode area, $\lambda$ is the laser vacuum wavelength, and $n_0$ and $n_2$ are the linear and nonlinear refractive indices respectively. The pump powers $|\tilde{f}_{\CW}|^2$ and $|\tilde{f}_{\CCW}|^2$ are normalised by $P_0/\eta_{\inn}$ where  $\eta_{\inn}=4\kappa(\gamma-\kappa)/\gamma^2$ is the in-coupling efficiency for coupling half-linewidth $\kappa$ and cavity half-linewidth $\gamma$.

We model a weak probe beam red-detuned by $\Omega$ from the pump frequency, while the pump beam itself is red-detuned by $\Delta$ from the cold cavity resonance, as seen in Fig.~1(d). Both $\Delta$ and $\Omega$ are dimensionless,  normalised by the coupled resonator half-linewidth $\gamma$. The fields can thus be expressed (in the frame co-rotating with the pump beam) as
\begin{equation}
\begin{split}
    \tilde{f}_{\CW} &= \alpha_0 \quad\quad
    \tilde{f}_{\CCW} = \alpha_1 e^{-i\Omega t}\\    
    f_{\CW} &= \beta_0 \quad\quad
    f_{\CCW} = \beta_1 e^{-i \Omega t},
    \end{split}
\end{equation}
where $\alpha_0$ and $\alpha_1$ are the amplitudes of the external pump and probe beams respectively. Setting the time derivatives of these amplitudes to zero ($
    \dot{\alpha}_0 =~ \dot{\alpha}_1 =~ \dot{\beta}_0 =~ \dot{\beta}_1 =~0$),
the intra-cavity amplitudes of the pump and probe beams, $\beta_0$ and $\beta_1$ respectively, satisfy
 \begin{equation}
    \begin{split}
     \beta_0 &= \frac{\alpha_0}{1 + i(-\Delta + |\beta_0|^2)}\\
     \beta_1 &= \frac{\alpha_1}{1+i(-\Delta - \Omega + 2 |\beta_0|^2)},
    \end{split}
 \end{equation}
where we neglected the Kerr shifts due to $\beta_1$ since $|\beta_0|~\gg~|\beta_1|$. The pump and probe beams experience SPM and XPM as seen by the factor of two difference in the Kerr shift. Thus, the scanning probe should measure a power-dependent resonance shift equivalent to the difference between the SPM and XPM. This is shown in Figs.~2(a) and 2(b), where an increase in the coupled pump power results in the scanning probe measuring the resonance further away from the pump frequency. 
From the intra-cavity fields, the pump and probe output fields are given~by
\begin{equation}
\label{alphaout}
    \begin{split}
        \alpha_{\out,0} &= \alpha_0 - 2\xi \beta_0 \\
        \alpha_{\out,1} &= \alpha_1 - 2\xi \beta_1,
    \end{split}
\end{equation}
where the fraction of light $\xi=\kappa/\gamma$ coupled out of the resonator can be found from $\eta_{\inn}$ by
\begin{equation}
    \xi = \frac{1 \pm \sqrt{1 - \eta_{\inn}}}{2},
    \label{xiITOetain}
\end{equation}
with the plus and minus signs corresponding to over- and under-coupling respectively. The intensities of the pump and probe outputs are detected by the photodiodes as $|\alpha_{\out,0}|^2$ and $|\alpha_{\out,1}|^2$,  respectively.

Probe-scanning traces were taken for different pump detunings from the SPM-shifted resonance, $\delta$, for a set of different coupled pump powers, as seen in Fig.~2(c). The in-coupled pump power is calculated by taking into account $\eta_{\inn}$, $\delta$, and $P_{\inn}$. The difference between XPM resonance and the pump frequency is plotted against coupled power in Fig.~2(c) together with the associated fitting error. In each measurement, $\delta$ is fixed by locking the transmitted power $P_\text{out}$ to a specific value that depends on maximum (out of resonance) and minimum (on resonance) transmission values, $P_\text{max}$ and $P_\text{min}$ respectively, as follows:
\begin{equation}
P_\text{out} = \frac{\delta^2P_\text{max}+\gamma^2P_\text{min}}{\delta^2+\gamma^2}.
\end{equation}
Each consecutive trace in Fig.~2(c) corresponds to a $\delta$ increase of 0.25$\gamma$, with $\gamma=2\pi\times$\SI{1.24}{\mega\hertz} obtained from the Lorentzian fits. The (common) slope $\gamma/P_0$ depends on both $\gamma$ and $\kappa$, obtained from $\eta_\text{in}~=~1~-~P_\text{min}~/~P_\text{max}$ via~(\ref{xiITOetain}), as well as the effective mode area $A_{\eff}$, calculated to be \SI{298}{\micro\meter^2}. Fig.~2(c) also illustrates the $\delta=0$ trace. Experimentally, cavity thermal instabilities prevented us from obtaining low-detuning data  \cite{carmon2004dynamical}.

\begin{figure}
 \centering\includegraphics[width=1\linewidth]{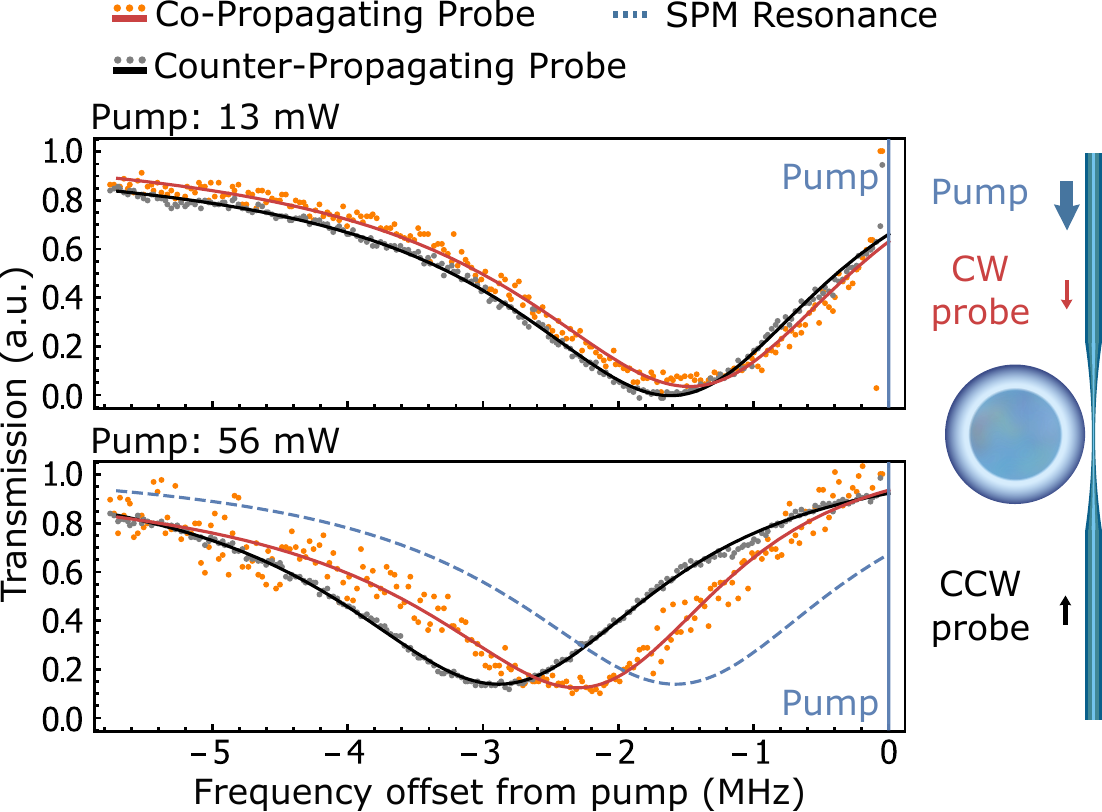}
 \caption{\label{fig:fig3}Measurement of XPM resonance shifts with two counterpropagating probes (red CW, black CCW), for both low and high pump powers, compared to the SPM-shifted pump resonance (dashed) to which the pump laser is locked (blue, CW). Interplay between XPM and FWM leads to a threefold splitting of the microresonator mode at high pump power.
 }
\end{figure}

For future applications, chip-based microresonators with lower nonlinear threshold powers like the microtoroid in Fig.~1(b) could be beneficial. Fig.~2(d) shows a measurement in a microtoroid with more than 35 cavity linewidths difference between XPM and SPM shift at 10~mW $P_\inn$, corresponding to \SI{286}{\micro\watt} of power for inducing a one cavity linewidth frequency splitting.

We were initially expecting that both the copropagating and counterpropagating resonances would undergo an identical shift in frequency, since XPM is independent of circulating direction. However, our probe spectroscopy measurements reveal an asymmetry in the shifts, such that the copropagating (CW) resonance undergoes less shift than originally expected. Moreover, an additional signal of the probe was detected mirrored around the pump frequency. The suggests the presence of FWM influencing the XPM-induced resonance shift. Degenerate FWM converts two pump photons into a probe and a signal photon, while the XPM-FWM interplay leads to a reduced resonance shift of the probe. The CW-CCW resonance shift difference is illustrated in Fig.~3. The resonator was probed simultaneously in both directions, suggesting that the pump interacted with the probe differently in the two directions, rather than the FWM decreasing the total power of the pump and hence the total shift. The data in Fig.~3 shows a threefold direction-dependent splitting of the cold cavity resonance.

\begin{figure}
\centering\includegraphics[width=1\linewidth]{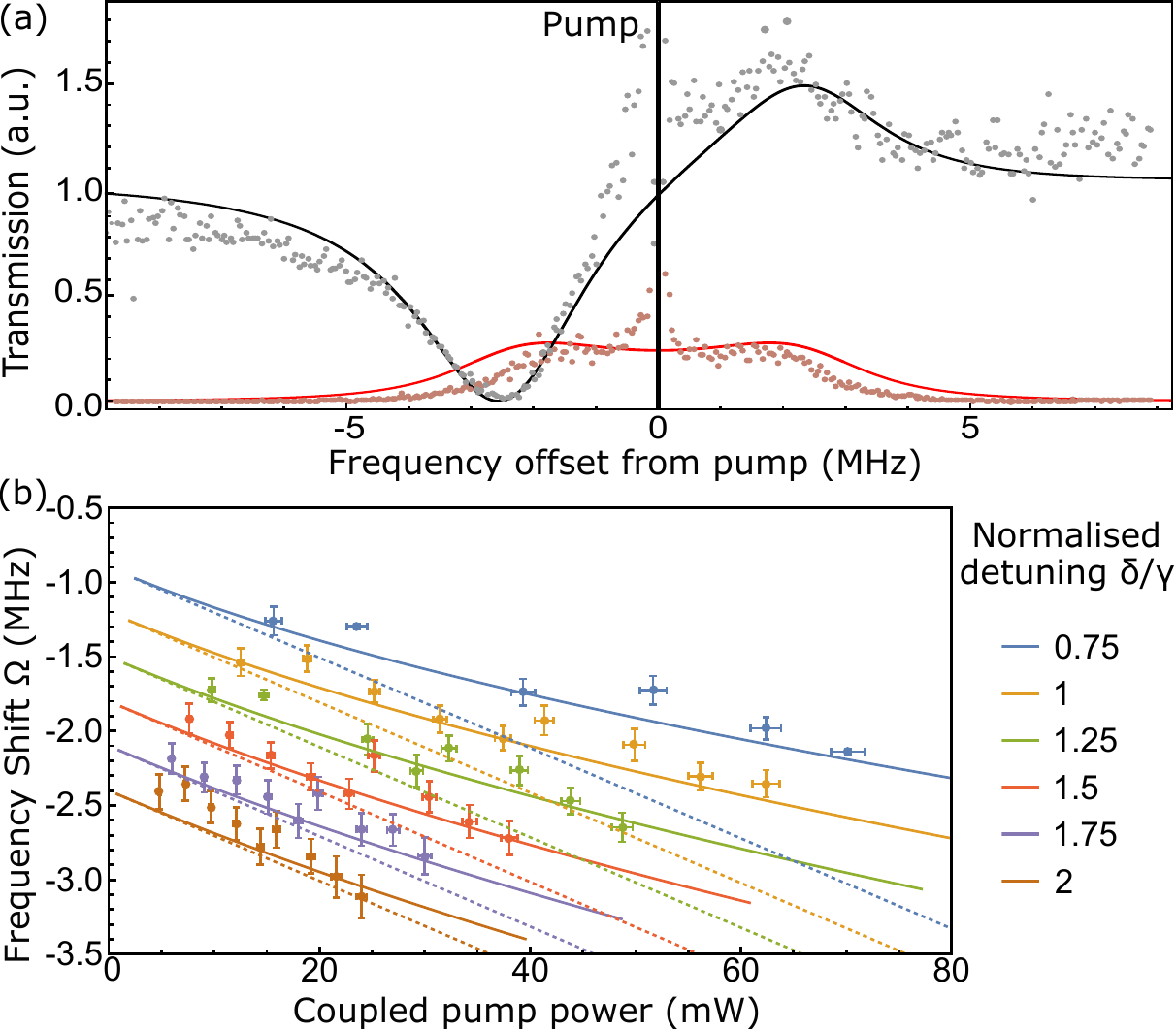}
\caption{\label{fig:fig4} (a) Transmitted probe power (black) and signal power (red) in the co-propagating pump-probe direction, with a simultaneous fit of $|\alpha_{\out, 1}|^2$ and $|\alpha_{\out, -1}|^2$ respectively. (b) Measured frequency difference between the pump and the centre of the co-propagating probe resonance vs.\ in-coupled pump power, for different values of $\delta/\gamma$. These data were taken simultaneously with the counter-propagating data shown in Fig.~2(c). The solid lines are theoretical curves based on the fitted parameters from Fig.~2. The dashed lines are the fitted shifts of the CCW resonance exactly as in Fig.~2(c).
}
\end{figure}

Our model is thus modified in the case of a CW probe that copropagates with the pump, to account for this FWM-XPM interplay. Similarly to the CCW probe case, we start from the dynamical equation
\begin{equation}
\label{eq:fDotCw}
    \dot{f}_{\CW} = \tilde{f}_{\CW} - (1-i\Delta)f_{\CW} - i|f_{\CW}|^2f_{\CW}
\end{equation}
that includes only clockwise-propagating light. We decompose the pump and circulating fields into frequency components as before, but additionally consider the presence of an intra-cavity `signal' field $\beta_{-1}$ resulting from FWM:
\begin{equation} 
\label{eq:fTildeCw}
\begin{split}
    \tilde{f}_{\CW} &= \alpha_0 + \alpha_1 e^{-i \Omega t} \\
    f_{\CW} &= \beta_0 + \beta_1 e^{-i \Omega t} + \beta_{-1} e^{i \Omega t},   
\end{split}
\end{equation}
Substituting equation (\ref{eq:fTildeCw}) into (\ref{eq:fDotCw}), and setting the time derivative of each component's amplitude to zero ($\dot{\alpha}_0~=~\dot{\alpha}_1~=~\dot{\beta}_0~=~\dot{\beta}_1~=~\dot{\beta}_{-1}~=~0$), we solve the equation such that it is true for all times $t$. Thus, we derive the amplitudes of the intra-cavity pump, probe, and signal beams respectively to be
\begin{align}
        \beta_0 &= \frac{\alpha_0}{1 + i(-\Delta + |\beta_0|^2)}\nonumber\\\nonumber\\
        \beta_1 &= \frac{\alpha_1(1 + i(\Delta - \Omega + 2|\beta_0|^2))}{(1 - i\Omega)^2 + (-\Delta + 2|\beta_0|^2)^2 - |\beta_0|^4}\\\nonumber\\\nonumber
        \beta_{-1} &= \frac{-i |\beta_0|^2 \alpha_1}{(1 - i\Omega)^2 + (-\Delta + 2|\beta_0|^2)^2 - |\beta_0|^4}\nonumber.
\end{align}
Similarly to the counterpropagating case, the output fields can be expressed as in (\ref{alphaout}), with the addition of the signal $
\alpha_{\out, -1} = -2\xi \beta_{-1}$,
emerging at the mirrored probe detuning $-\Omega$ from the pump.

By plotting the output powers $|\alpha_{\out, 1}|^2$ and $|\alpha_{\out, -1}|^2$ with respect to the detuning $\Omega$, we show two very distinct characteristics of the spectrum, as shown in Fig.~4(a): first, the model predicts the generation of two peaks in the power of the signal beam, which occur when either the probe or signal coincides in frequency with the XPM-and-FWM-shifted resonance. 
The second characteristic feature of FWM predicted by our model is the small peak in the probe trace, seen in Fig.~4(a). 

As illustrated in Fig.~4(b), our model shows the power-dependent frequency shift, for the same parameters used earlier to fit the counter-propagating data. Note that the CW and CCW probe measurements were taken simultaneously, and therefore correspond to the same resonance, coupling, and pump power. For comparison, the dotted lines in Fig.~4(b) correspond to the theoretical prediction without FWM for the CCW case. While the CW data exhibit bigger error bars due to pump noise in the frequency range of the measurement (since, unlike in the CCW case, the intense pump is incident on the same photodiode as the probe), our model still fits these data using the exact same parameters. 

In conclusion, we have demonstrated resonance frequency splittings induced by XPM for continuous wave lasers in ultra-high-Q microresonators. The mode splittings are directly measured with a pump-probe spectroscopy scheme. In addition, we show that the XPM-FWM interplay can lead to a threefold resonance splitting. Using a microtoroid resonator, we show Kerr splittings of one cavity linewidth at very low powers of only \SI{286}{\micro\watt}. Resonance splitting of up to 35 cavity linewidths is observed at 10 mW optical power. With optical microresonators being building blocks for photonic integrated circuits, precise and fast control over their resonances is of critical importance. The optically induced splitting of resonances through the Kerr-nonlinearity at low power levels could enable new devices e.g. for coherent coupling of light to phonon modes that match the optical frequency splitting \cite{schliesser2008resolved, vahala2009phonon}. More applications include optical memories and the use of microresonators for optically controlled modulators and filters.
\\

Funded by H2020 European Research Council (ERC) (756966,
CounterLight); H2020 Marie Sklodowska-Curie Actions
(MSCA) (CoLiDR, 748519; GA-2015-713694);
Engineering and Physical Sciences Research Council
(EPSRC) (CDT for Controlled Quantum Dynamics,
Applied Photonics, and Quantum Systems Engineering
Skills Hub); Aker Scholarship; National Physical
Laboratory (NPL), Max Planck Institute for the Science
of Light (MPL).

\bibliographystyle{apsrev4-1}
\normalem
\bibliography{mybib}

\end{document}